\documentclass[aps,superscriptaddress,nofootinbib,showpacs]{revtex4}
\usepackage{epsfig,rotating}
\usepackage{amsmath,amssymb}
\usepackage{dsfont}

\newcommand{\be}{\begin{equation}}
\newcommand{\ee}{\end{equation}}
\newcommand{\bea}{\begin{eqnarray}}
\newcommand{\eea}{\end{eqnarray}}

\newcommand{\vk}{\vec{k}}

\newcommand{\om}{\omega}

\newcommand{\ga}{\gamma}

\newcommand{\uvk}{\widehat{\bf{k}}}

\begin{document}

\title{Charged lepton mixing and oscillations from neutrino mixing in the early
Universe.}
\author{\bf D. Boyanovsky }
\email{boyan@pitt.edu}\affiliation{  Department of Physics and
Astronomy, University of Pittsburgh, Pittsburgh, Pennsylvania 15260,
USA }
\author{\bf C. M. Ho}
\email{cmho@phyast.pitt.edu}\affiliation{  Department of Physics and
Astronomy, University of Pittsburgh, Pittsburgh, Pennsylvania 15260,
USA }

\date{\today}

\begin{abstract}
Charged lepton mixing   as a consequence of neutrino mixing is
studied for two generations $e,\mu$ in the temperature regime $m_\mu
\ll T \ll M_W$ in the early Universe. We state the general criteria
for charged lepton mixing, critically reexamine aspects of neutrino
equilibration and provide arguments to suggest that neutrinos may
equilibrate as mass eigenstates in the temperature regime
\emph{prior} to flavor equalization. We assume this to be the case,
and that neutrino mass eigenstates are in
  equilibrium with different chemical potentials. Charged
lepton self-energies are obtained to leading order in the
electromagnetic and weak interactions. The upper bounds on the
neutrino asymmetry parameters from CMB and BBN without oscillations,
combined with the fit to the solar and KamLAND data for the neutrino
mixing angle, suggest that for the two generation case there is
resonant \emph{charged lepton} mixing in the temperature range $T
\sim 5  \,\mathrm{GeV}$.  In this range the charged lepton
oscillation frequency is of the same order as the electromagnetic
damping rate.
\end{abstract}


 \maketitle

Neutrinos  play a fundamental role in cosmology and astrophysics
\cite{raffelt}, and there is now indisputable  experimental
confirmation that neutrinos are massive and that different flavors
of neutrinos mix and oscillate \cite{panta,giunti,smirnov, haxton,
grimus,gouvea}, thus providing evidence for \emph{new physics}
beyond the Standard Model. Neutrino oscillations in extreme
conditions of temperature and density are an important aspect of Big
Bang Nucleosynthesis (BBN), in the generation of the lepton
asymmetry in the early
Universe\cite{fuller,dolgov,haxton,raffelt,kirilova}, and in the
physics of core collapse supernovae\cite{SN,panta}.

An important aspect of neutrino oscillations is lepton number
violation, leading to the  suggestion that  leptogenesis can be a
main ingredient in an  explanation of the cosmological baryon
asymmetry \cite{fukugita}. Early studies of neutrino propagation in
hot and dense media  focused on the neutrino dispersion relations
and damping rates in the temperature regime relevant for stellar
evolution or big bang nucleosynthesis\cite{notzold,dolgov,raffelt}.
This work has been extended to include leptons, neutrinos and
nucleons in the medium \cite{dolivoDR}. Matter effects of neutrino
oscillations in the early universe were investigated in
\cite{barbieri,enqvist} and more recently a field theoretical
description of mixing and oscillations in real time has been
provided in ref.\cite{hoboya}. While there is a large body of work
on the study of neutrino mixing in hot and dense environments, much
less attention has been given to the possibility of mixing and
oscillation of \emph{charged leptons}. Charged lepton number
non-conserving processes, such as $\mu \rightarrow e\,\gamma; \mu
\rightarrow 3\, e$ mediated by massive mixed neutrinos have been
studied in the vacuum in refs.\cite{lee,petcov,bile}. For Dirac
neutrinos the transition probabilities for these processes are
suppressed by a factor $m^4_a/M^4_W$\cite{lee,petcov,bile}. The
WMAP\cite{WMAP} bound on the neutrino masses $m_a < 1\, \mathrm{eV}$
yields typical branching ratios for these processes $B \lesssim
10^{-41}$ making them all but experimentally unobservable. In this
article we explore the possibility of \emph{charged lepton mixing}
in the early Universe at high temperature and density. In section
 \ref{mixin}  we discuss the general arguments for charged lepton
mixing as a result of neutrino mixing and establish the necessary
conditions for this mixing to be substantial. We suggest that large
neutrino chemical potentials may lead to substantial charged lepton
mixing.

Without oscillations BBN and CMB provide a stringent constraint on
the neutrino chemical potentials\cite{hansen,kneller} $\xi_\alpha$,
with $-0.01\leq \xi_e \leq 0.22\,,\,|\xi_{\mu,\tau}|\leq 2.6$.
Detailed studies\cite{fuller,luna,doleq,wong,aba} show that
oscillations and self-synchronization lead to flavor equilibration
before BBN, beginning at a temperature $T \sim
30\,\mathrm{MeV}$\cite{doleq} with complete flavor equilibration
among the chemical potentials at $T \sim
2\,\mathrm{MeV}$\cite{luna,doleq}. Thus \emph{prior to flavor
equalization} for $T  > 30\,\mathrm{MeV}$ there \emph{could} be
large neutrino  asymmetries consistent with the BBN and CMB bounds
in the absence of oscillations. We study whether this possibility
could lead to  charged lepton  mixing focusing on two flavors of
Dirac neutrinos corresponding to $e,\mu$ and for simplicity in the
temperature regime where both are ultrarelativistic, with $m_\mu \ll
T\ll M_W$. In section  \ref{equil}  we discuss   general arguments
within the realm of reliability of perturbation theory   suggesting
that the equilibrium state is described by a density matrix nearly
diagonal in the mass basis. In this section we also discuss caveats
and subtleties in the kinetic approach to neutrino equilibration in
the literature and argue that results on the equilibrium state are
in agreement with the interpretation of an equilibrium density
matrix diagonal in the mass basis. Our \emph{main and only
assumption} is that for $T> 30\, \textrm{MeV}$ neutrinos are in
equilibrium and the density matrix is nearly diagonal in the mass
basis, with distribution functions of mass eigenstates that feature
different and large chemical potentials. While this is not the
\emph{only}, it is \emph{one} possible scenario for  substantial
charged lepton mixing that can be explored systematically.   In
section
 \ref{mixing}  we explore charged lepton mixing in lowest order in
perturbation theory as a consequence of large asymmetries in the
equilibrium distribution functions of mass eigenstates. In this
section we also critically discuss possible caveats and suggest a
program to include non-perturbative corrections in a systematic
expansion. In section \ref{conclu} we summarize the main aspects and
results of the article.

\section{Charged lepton mixing: the general argument}\label{mixin}

\begin{figure}[h]
\begin{center}
\includegraphics[height=3in,width=4in,keepaspectratio=true]{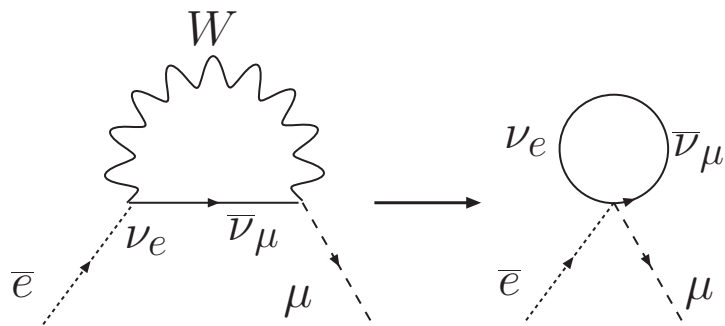}
\caption{Off diagonal charged lepton self energy: (a) one loop
$W$-boson exchange, (b)  self-energy in the effective Fermi theory.}
\label{loopmixing}
\end{center}
\end{figure}

Charged lepton mixing is a consequence of neutrino mixing in the
charged current contribution to the charged lepton self energy. This
can be seen as follows: consider the one-loop self energy for the
charged leptons. The off-diagonal self-energy $\Sigma_{e\mu}$ is
depicted in fig. (\ref{loopmixing}) for the case of   electron-muon
mixing. The internal line in fig. (\ref{loopmixing}) (a) is a
neutrino propagator off-diagonal in the flavor basis, which is
non-vanishing if neutrinos mix. In Fermi's effective field theory
obtained by integrating out the vector bosons the effective
interaction that gives rise to charged lepton mixing is

\be H_{eff}= \frac{2G_F}{\sqrt{2}} \left[\overline{e}_L \gamma_\mu
\nu_{eL}\right]\left[ \overline{\nu}_{\mu L}\gamma_\mu \mu_L
\right]\,. \label{Heff} \ee

A simple Hartree-like factorization yields

\be \frac{2G_F}{\sqrt{2}}  \overline{e}_L \gamma_\mu < \nu_{eL}
\overline{\nu}_{\mu L}> \gamma^\mu \mu_L \equiv \overline{e}_L\,
\Sigma_{e\mu}\, \mu_L \label{selfe}\ee  where the brackets stand for
average in the density matrix of the system. Eqn. (\ref{selfe})
gives the charged-lepton mixing part of the self energy as

\be \Sigma_{e\mu} = \frac{2G_F}{\sqrt{2}}\,\gamma^\mu< \nu_{eL}
\overline{\nu}_{\mu L}>\gamma_\mu \label{ave} \,.\ee The Fermi
effective field theory contribution to the self-energy is depicted
in fig. (\ref{loopmixing}) (b).

\vspace{1mm}

The focus of this article is to study two aspects that emerge from
this observation:

\begin{itemize}
\item{{\bf Mixing:} The propagating modes in the medium are determined by the
poles of the full propagator with a self-energy that includes
radiative corrections in the medium. The full self-energy for the
charged leptons is a $2\times 2$ matrix (in the simple case of two
flavors), and eqn. (\ref{ave}) yields the off-diagonal matrix
element in the flavor basis.  This is precisely the main study in
this article: we obtain the charged lepton propagator including
radiative corrections in the medium up to one loop in the
electromagnetic and weak interactions. Neutrino mixing leads to off
diagonal components of the propagator in the charged lepton flavor
basis. We find the dispersion relation of the true propagating modes
in the medium by diagonalization of the \emph{full} propagator
including one loop radiative corrections. The true propagating modes
in the medium are \emph{admixtures} of electron and muon states:
this  is precisely what we identify as \emph{mixing}. The results
given by equations (\ref{selfe},\ref{ave}) state \emph{quite
generally} that electron and muon states are mixed whenever
\emph{the neutrino propagator is off-diagonal in the flavor basis}.
We highlight that this is \emph{precisely} the condition for
\emph{flavor neutrino oscillations} since the propagator yields the
transition amplitude from an initial to a final state. Therefore we
state quite generally that provided flavor neutrinos oscillate,
namely if the neutrino propagator is \emph{off diagonal in the
flavor basis}, charged leptons associated with these flavor
neutrinos will mix. The true propagating modes of charged leptons
are linear superpositions of the charged leptons associated with the
flavor neutrinos. We emphasize these statements because even though
they are a straightforward consequence of flavor neutrino mixing,
this precise point, and its consequences,  have not been previously
addressed in the literature. }

\item{{\bf Oscillations:} Consider the decays $W\rightarrow \nu_e\,e$ or neutron
beta decay $n\rightarrow p\,e\,\overline{\nu}$ in the medium. The
electron produced in the medium at the decay vertex propagates
\emph{as a linear combination} of the true propagating modes in the
medium, each with a different dispersion relation.  Upon time
evolution this linear superposition will have non-vanishing overlap
with a muon state yielding a typical oscillation pattern. We study
this \emph{oscillation} between the electron and muon charged lepton
by considering the evolution of an electron wave-packet produced
locally at the decay vertex. These oscillations are akin to the
typical oscillation between flavor neutrino states and are a
consequence of the one loop radiative correction depicted in
fig.(\ref{loopmixing}) with neutrinos in the medium. The transition
probability from an initial electron to a muon packet oscillates in
time. While in the case of almost degenerate neutrinos oscillations
are associated with \emph{macroscopic quantum coherence} because the
oscillation lengths are macroscopically large, this is not a
necessary condition for oscillations, which occur whenever the
initial state is a linear superposition of the propagating modes.
The example of neutron beta decay gives a precise meaning to the
statement of charged lepton mixing: in the decay of the neutron the
charged lepton that is produced is identified with the electron.
This is the initial state, which in a medium will propagate as a
linear combination of the propagating modes with an oscillatory
probability of finding a muon. These oscillations are
\emph{fundamentally different} from the space-time oscillations
possibly associated with quantum entanglement and discussed in
references\cite{entanglement}. }

\end{itemize}

Eqn. (\ref{ave}) generally  states that there is charged lepton
mixing   when the density matrix is \emph{off diagonal} in the
flavor basis. This   is equivalent to the statement of neutrino
mixing. A simple example of a density matrix off diagonal in the
flavor basis is $\hat{\rho} = |0_m><0_m|$ with $|0_m>$ being the
vacuum state in absence of weak interactions but with a neutrino
Hamiltonian with an off diagonal mass matrix in the flavor basis.
This is the interaction picture vacuum of the standard model
augmented by a neutrino mass matrix with flavor mixing. In the two
flavor case with \be \nu_e(\vec{x},t) = \cos\theta \,\nu_1
(\vec{x},t)+ \sin \theta \, \nu_2(\vec{x},t) \;,\;\nu_\mu
(\vec{x},t)= \cos\theta \,\nu_2(\vec{x},t) -\sin \theta \,\nu_1
(\vec{x},t) \label{twoflavmix} \ee with $\nu_{1,2}(\vec{x},t)$ the
fields associated with the mass eigenstates, \be <0_m|
\nu_{e}(\vec{x},t) \overline{\nu}_{\mu}(\vec{x}',t')|0_m> =
\cos\theta \sin \theta \left[ < 0_m|\nu_{2}(\vec{x},t)
\overline{\nu}_{2}(\vec{x}',t')|0_m>- <0_m| \nu_{1}(\vec{x},t)
\overline{\nu}_{1}(\vec{x}',t')|0_m>\right]\,.\label{diff} \ee If
the propagators for the mass eigenstates only differ in the masses,
this difference leads to a very small self-energy. In a medium the
flavor off diagonal expectation value (\ref{ave}) could be enhanced
by temperature and or density. Therefore the \emph{general}
criterion for  substantial  charged lepton mixing in a medium hinges
on just one aspect: a large off diagonal matrix element $< \nu_{eL}
\overline{\nu}_{\mu L}>$. One possible case for which this condition
is fulfilled is \emph{if} the density matrix is nearly diagonal in
the \emph{mass basis} with large chemical potentials for the
different mass eigenstates.

We emphasize that   this is only \emph{one} condition for
substantial charged lepton mixing and by no means unique, the
analysis above   shows that the most general condition is simply
that $< \nu_{eL} \overline{\nu}_{\mu L}>$ be large.

In the general case a full solution of a kinetic equation should
yield the value of $< \nu_{eL} \overline{\nu}_{\mu L}>$. \emph{If}
an equilibrium state of mixed neutrinos is described by a density
matrix   nearly diagonal in the mass basis with distribution
functions for the different mass eigenstates with large and
different chemical potentials, then simple expressions for the
equilibrium propagators allow an assessment of the charged lepton
mixing self energy. Can this be the case?.

\section{On neutrino equilibration}\label{equil}
\subsection{Equilibration in the mass basis}\label{massbasis}

A system is in equilibrium if \be \left[\hat{\rho},H\right] =0 \,,
\label{commute}\ee  where $\hat{\rho}$ is the density matrix of the
system and $H$ the total Hamiltonian $H=H_0+H_{int}$ with $H_0$ the
Hamiltonian in the absence of weak interactions but with a mass
matrix and $H_{int}=H_{NC}+H_{CC}$. In the absence of weak
interactions, an equilibrium density matrix $\hat{\rho}_0$ commutes
with $H_0$, therefore it is diagonal in the \emph{mass basis}. The
equilibrium density matrix cannot have off-diagonal matrix elements
in the mass basis because these oscillate in time. Of course without
interactions the system will not reach an equilibrium state,
however, as is the usual assumption in statistical mechanics,
provided the interactions are sufficiently weak but lead to an
equilibrium state, an \emph{almost} free gas of particles in
equilibrium is a suitable description, and the canonical density
matrix in such case is of the form $\hat{\rho}_0 = e^{-H_0/T}$ (the
grand canonical could also include a chemical potential for
conserved quantities). Examples of this are abundant, a ubiquitous
one is the cosmic microwave background radiation: the Planck
distribution function describes free photons in equilibrium,
although  photons reach equilibrium by undergoing  collisions with
charged particles in a plasma with cross sections much larger than
those of neutrinos.

Consider   how the density matrix is modified from the ``free
field'' form by  ``switching on'' the  weak interactions in
perturbation theory. A perturbative expansion in the interaction
picture of $H_0$ begins by writing the   interaction vertices in
terms of neutrino fields in the \emph{mass basis}. Neglecting
sterile neutrinos,   neutral current interaction vertices are
diagonal   and only the charge current interactions induce
off-diagonal correlations in the mass basis. Let us write the full
density matrix as $\hat{\rho}=\hat{\rho}_0+\delta{\hat{\rho}}$ where
$\delta{\hat{\rho}}$ has a perturbative expansion in the weak
coupling. The equilibrium condition leads to the following identity

\be \left[\delta{\hat{\rho}},H_0 \right] = -
\left[\hat{\rho}_0,H_{int} \right]- \left[\delta{\hat{\rho}},H_{int}
\right]\label{equilint}\ee  Taking matrix elements in the mass
eigenstates of $H_0$ the solution of eqn. (\ref{equilint})  for the
  matrix elements of $\delta{\hat{\rho}}$ in the mass basis can be found in a
perturbative expansion. The matrix elements of $\delta{\hat{\rho}}$
may feature non-vanishing off diagonal correlations in the mass
basis as a result of  charged current vertices which mix different
mass eigenstates. However, a perturbative solution for the matrix
elements of eqn. (\ref{equilint}) in the mass basis would at most
result in off diagonal correlations which are \emph{perturbatively}
small. Namely, the equilibrium density matrix is nearly diagonal in
the mass basis.

Expanding the field operators associated with the mass eigenstates
in terms of Fock creation and annihilation operators of mass
eigenstates, the spatial Fourier transform of the field operators is
given by  \be \nu_{i}(\vec{k},0) = \sum_{\lambda}
a_i(\vec{k},\lambda)\,\mathcal{U}_i(\vec{k},\lambda)+
b^{\dagger}_i(-\vec{k},\lambda)\,\mathcal{V}_i(-\vec{k},\lambda)~~;~~
i=1,2 \ee where the spinors $\mathcal{U},\mathcal{V}$ are
orthonormalized positive and negative energy solutions solutions of
the Dirac equation with mass $m_i$.  If the density matrix is
diagonal in the mass basis, then $\langle
a^{\dagger}_i(\vec{k})\,a_j (\vec{k})\rangle \propto \delta_{ij}$
and the distribution functions for the mass eigenstates are $\langle
a^{\dagger}_i(\vec{k})\,a_i(\vec{k}) \rangle\,;\,\langle
b^{\dagger}_i(\vec{k})\,b_i(\vec{k}) \rangle$ for neutrinos and
antineutrinos of mass $i$ respectively. Switching on the
\emph{neutral current interaction} which is diagonal in the mass
basis (provided there are no sterile neutrinos) will lead to the
equilibration of neutrinos and the equilibrium distribution
functions will be the usual Fermi-Dirac with a possible chemical
potential. The \emph{charged current interactions} yield vertices
that are off-diagonal in the mass basis and   induce cross
correlations of the form $\langle a^{\dagger}_i\,a_j \rangle$ with
$i\neq j$. In free field theory this equal time correlation
function, if non-vanishing, oscillates with a time dependence $e^{i
(\omega^{i}_k-\omega^{j}_k) \,t}$, however, in  equilibrium there
cannot be a time dependence of these off diagonal correlations as
the following argument shows \be
  \langle a^{\dagger}_i(t)\,a_j(t) \rangle  =
\mathrm{Tr}\left( \hat{\rho}e^{iHt}  a^{\dagger}_i(0)\,a_j(0)
  e^{-iHt}\right)= \mathrm{Tr} \left( \hat{\rho}
a^{\dagger}_i(0)\,a_j(0) \right) \ee where we used eqn.
(\ref{commute}). Either the charged current interactions that
generate these off-diagonal correlations \emph{exactly} cancel the
free field  time dependence for all values of momentum $k$ or, more
likely, they lead to the decay of these off diagonal correlations to
asymptotically perturbatively small expectation values as expected
from the general arguments following eqn. (\ref{equilint}).

 This observation leads to
the conclusion that \emph{if} the perturbative expansion is
reliable, the weak interactions lead to an equilibrium state
described by a density matrix which is nearly diagonal in the mass
basis but for possible perturbatively small off-diagonal elements.
In perturbation theory the \emph{equilibrium} distribution functions
are diagonal in the mass basis and may feature a chemical potential
for each mass eigenstate. Weak interaction vertices involve the
flavor fields, but these are linear combinations of the fields that
create and annihilate mass eigenstates, the true in-out states.
Consider a far off-equilibrium initial state with a population of
vector bosons (or neutrons) and no neutrinos, the decay of the
vector bosons (or neutrons) results in the creation of a linear
superposition of mass eigenstates, which propagate independently
after production. Collisional processes via the weak interaction
lead to the decoherence of the mass eigenstates and ultimately to a
state of \emph{equilibrium} in which equal time expectation values
in the density matrix cannot depend on time. Neutral and charged
current interactions yield different relaxational dynamics: in the
mass basis the neutral current interaction  is diagonal and
relaxation processes via neutral currents lead to equilibration in
the mass basis. Charged currents feature both diagonal and
off-diagonal contributions in the mass basis, the diagonal ones
yield relaxation dynamics similar to the neutral current
interaction. The off diagonal contributions induce correlations
between different mass eigenstates, but also lead to the relaxation
of these off diagonal correlations. These two types of processes
leading to relaxation dynamics for diagonal and off-diagonal
correlations in the mass basis are akin to the different processes
that lead to the relaxation times $T_1$ (diagonal) and $T_2$
(transverse) in spin systems in nuclear magnetic
resonance\cite{slichter}. These concepts are manifest in Stodolsky's
effective Bloch equation description of neutrino oscillations in  a
medium with a damping coefficient in the ``transverse''
direction\cite{stod} whose inverse is the equivalent of the $T_2$
relaxation time in spin systems.

By the above arguments this asymptotic equilibrium density matrix
must be nearly diagonal in the mass basis at least within the realm
of reliability of perturbation theory. Equilibrium correlation
functions of operators at different times must be functions of the
time difference. Of particular relevance to the   discussion below
is the flavor off diagonal propagator $\langle
 {\nu}_e(\vk,t) \overline{\nu}_{\mu}(\vk,t')\rangle$. Writing the
flavor fields as linear combinations of the fields $\nu_{1,2}$ this
correlation function in the equilibrium density matrix diagonal in
the mass basis is to zeroth order in the perturbation \be \langle
 {\nu}_e(\vk,t)\, \overline{\nu}_{\mu}(\vk,t')\rangle =
-\cos\theta \sin\theta\left[\langle  {\nu}_1(\vk,t-t')
\,\overline{\nu}_{1}(\vk,0)\rangle -\langle
 {\nu}_2(\vk,t-t')\, \overline{\nu}_{2}(\vk,0)\rangle\right]\,.
\label{offlavor}\ee Mixed correlators of $\nu_{1,2}$ \emph{cannot}
be functions of the time difference because of the different masses
lead to a dependence on $t+t'$. As a simple but relevant example,
consider the density matrix for the \emph{vacuum} \be \hat{\rho} =
|0><0| \ee where $|0>$ is the exact ground state of H. This state
can be constructed
 systematically  in perturbation theory from the ground state $|0_m>$ of the
Hamiltonian $H_0$ in absence of weak interactions, namely the
interaction picture ground state in the basis of mass eigenstates,
\be |0> = |0_m>+\sum_n |n_m>\frac{<n_m|H_{int}|0_m>}{-E_n}+\cdots\,.
\ee where $|n_m>$ are Fock eigenstates of $H_0$ (``mass
eigenstates'') with energy $E_n$. Writing the full density matrix as
$\hat{\rho} = |0_m><0_m| +\delta{\hat{\rho}}$ one can find
$\delta\hat{\rho}$ systematically in perturbation theory. The
off-diagonal flavor propagator \bea S_{e\mu}(\vk,t-t') & = & <0|
{\nu}_e(\vk,t)\overline{\nu}_{\mu}(\vk,t')|0> = <0_m|
{\nu}_e(\vk,t)\overline{\nu}_{\mu}(\vk,t')|0_m>+
\mathcal{O}(g)+\cdots \nonumber \\ & = &  \cos \theta \sin\theta
\left[<0_m| {\nu}_2(\vk,t)\overline{\nu}_{2}(\vk,t')|0_m> -<0_m|
{\nu}_1(\vk,t)\overline{\nu}_{1}(\vk,t')|0_m>\right] +
\mathcal{O}(g)+\cdots\,. \label{propavac} \eea This propagator is
the lowest order intermediate state in the process $W\,e \rightarrow
W\,\mu$, and is also the internal fermion line, along with W-vector
boson exchange in the off-diagonal self-energy contribution for
charged leptons, see fig.(\ref{loopmixing}). This simple example
also leads to conclude that if there is an equilibrium state for
which the equal time ``distribution function'' $<
{\nu}_e(\vk)\overline{\nu}_\mu(\vk)> \neq 0$ then the \emph{unequal}
time correlation function \be \label{propamix}
\mathcal{S}_{e\mu}(\vk, t-t') = <
{\nu}_e(\vk,t)\overline{\nu}_\mu(\vk,t')> = \cos\theta \sin\theta
\left[< {\nu}_2(\vk,t)\overline{\nu}_2(\vk,t')> -<
{\nu}_1(\vk,t)\overline{\nu}_1(\vk,t')>  \right] \neq 0\ee As it
will become clear below, this is the correlation function that
describes the charged lepton mixing. These arguments rely on the
validity of the perturbative expansion and require revision in the
case where perturbation theory in the mass basis must be reassessed.
In section (\ref{PT}) we discuss this possibility and propose a
method to re-arrange the perturbative expansion.

\subsection{ On the kinetic approach}\label{kinetics}

 Early kinetic approaches to
the dynamics of oscillating neutrinos in thermal environments were
proposed by Dolgov\cite{dol,dolgov,barbieri}, Stodolsky\cite{stod}
and Manohar\cite{mano}. Dolgov\cite{dol} introduced density matrices
in flavor space, whereas Stodolsky\cite{stod} and Manohar\cite{mano}
used a single particle density matrix of flavor states leading to
Bloch-like equations and a similar description was studied in
ref.\cite{enqvist}. Stodolsky argued that decoherence between flavor
states emerged from a ``transverse'' relaxation akin to the
relaxation time $T_2$ in nuclear magnetic resonance\cite{slichter}.
A Boltzmann equation for mixing and decoherence was established by
Raffelt, Sigl and Stodolsky\cite{rafstod} in terms of a ``matrix of
densities'' in the \emph{nonrelativistic domain} instead of the
density matrix. In this approach the field operators for different
flavors were truncated to only the annihilation operators and
obtained a Boltzmann equation in a perturbative expansion. A fully
relativistic treatment was presented in ref.\cite{rafsigl}
introducing ``matrices of densities'' defined by the expectation
value of bilinears of creation and annihilation operators of flavor
states. A quantum kinetic description of oscillating neutrinos was
presented in ref.\cite{mckellar} with an approach similar to those
of refs.\cite{stod,rafstod} in terms of single particle flavor
states.

All of the approaches to the kinetic description of oscillating
neutrinos in a medium in one way or another use the notion of flavor
Fock states, either in terms of single particle flavor neutrino
states or by expanding field operators in terms of creation and
annihilation Fock operators for flavor states. However, flavor
states and the precise definition of  Fock operators associated with
these states are very subtle and
controversial\cite{giunti,blasone,fujii,field,ji,li}.

\vspace{2mm}

{\bf Precisely, what are flavor states?} While in the literature
there is no precise definition of a ``flavor state'', a proper
definition of such state should begin by expanding the flavor field
operator in terms of Fock creation and annihilation operators. In
such an expansion the spatial Fourier transform of the flavor
neutrino field operator, for example the electron neutrino at $t=0$,
is given by \be \label{flavorfield} \nu_e(\vec{k},0) =
\sum_{s}\alpha_e(\vec{k},s)\mathcal{U}_e(\vec{k},s)+\beta^{\dagger}_e(-\vec{k},s)\mathcal{V}_e(-\vec{k},s)\ee
where the spinors $\mathcal{U}_e\,;\,\mathcal{V}_e$ are
orthonormalized positive and negative frequency solutions of a Dirac
operator with \emph{some mass} and define a basis. A Fock state of
an electron neutrino \emph{can} be defined by \be
|\nu_e(\vec{k},s)\rangle =
\alpha^{\dagger}_e(\vec{k},s)|0_m\rangle\ee where $ |0_m\rangle$ is
the   vacuum of the non-interacting theory, namely the vacuum of
mass eigenstates. However, the expansion (\ref{flavorfield})
requires a definite basis corresponding to a definite choice of the
Dirac spinors $\mathcal{U},\mathcal{V}$, which can be chosen to be
solutions of a Dirac operator for any arbitrary mass. Each possible
choice of mass gives a different definition of ``particle''. One
possible choice is zero mass\cite{rafsigl}, another choice would be
the diagonal elements of the mass matrix in the flavor
basis\cite{hoboydense} or masses $m_1$ to the electron neutrino and
$m_2$ to the muon neutrino\cite{blasone}. Any of these choices is
just as good and physically motivated but obviously arbitrary. The
creation and annihilation operators are extracted by
projection\cite{hoboydense}, for example
$\alpha^{\dagger}_e(\vec{k},s) = \nu^{\dagger}_e(\vec{k},0)
\mathcal{U}_e(\vec{k},s)$. Writing the electron neutrino field
operator as a linear combination of the field operators that create
and annihilate mass eigenstates one finds \bea
\alpha^{\dagger}_e(\vec{k},s)  = && \cos \theta \left[\sum_{\lambda}
a^{\dagger}_1(\vec{k},\lambda)\,\mathcal{U}^{\dagger}_1(\vec{k},\lambda)\mathcal{U}_e(\vec{k},s)+
b_1(-\vec{k},\lambda)\,\mathcal{V}^{\dagger}_1(-\vec{k},\lambda)\mathcal{U}_e(\vec{k},s)
\right] + \nonumber \\ && \sin \theta  \left[\sum_{\lambda}
a^{\dagger}_2(\vec{k},\lambda)\,\mathcal{U}^{\dagger}_2(\vec{k},\lambda)\mathcal{U}_e(\vec{k},s)+
b
_2(-\vec{k},\lambda)\,\mathcal{V}^{\dagger}_2(-\vec{k},\lambda)\mathcal{U}_e(\vec{k},s)
\right] \eea

The transformation between the set of ``flavor'' operators and those
that create and annihilate mass eigenstates is unitary and the
scalar products of the spinors yield  generalized Bogoliubov
coefficients.  It is clear that there is no single choice of spinor
$\mathcal{U}_e$ that will make \bea \label{overlapsUV}&&
\mathcal{U}^{\dagger}_1(\vec{k},\lambda)\mathcal{U}_e(\vec{k},s) =1
\, ; \,
\mathcal{U}^{\dagger}_2(\vec{k},\lambda)\mathcal{U}_e(\vec{k},s) =1
\nonumber \\ &&
\mathcal{V}^{\dagger}_1(-\vec{k},\lambda)\mathcal{U}_e(\vec{k},s)=0
\, ; \,
\mathcal{V}^{\dagger}_2(-\vec{k},\lambda)\mathcal{U}_e(\vec{k},s)=0
\eea

A surprising result of the above identification is that the
annihilation operator $\alpha_e(\vec{k},s)$ \emph{creates a linear
combination of antineutrino mass eigenstates} out of the vacuum
$|0_m>$\cite{hoboydense,blasone,fujii,ji,li}.

This  observation indicates that \emph{any} choice of the solutions
for the spinors $\mathcal{U}_e,\mathcal{V}_e$ to define a flavor
Fock creation operator leads to $|\nu_e \rangle \neq \cos \theta
|\nu_1\rangle + \sin\theta |\nu_2\rangle$. A possible definition of
flavor Fock states would be to \emph{define} the flavor vacuum
$|0_f>$ as the state annihilated by the flavor annihilation
operators (defined for example for zero mass states) and to
construct a Fock Hilbert space out of this vacuum by successive
application of   flavor Fock creation operators. However, while
there is a formal unitary transformation that relates the flavor and
mass Fock operators via the Bogoliubov coefficients, such
transformation is \emph{not} unitarily implementable in the infinite
dimensional Hilbert space\cite{blasone}, in particular
$<0_f|0_m>=0$. Finally one can simply \emph{define} flavor states as
\be \label{deflavor} |\nu_e> \equiv \cos\theta
|\nu_1>+\sin\theta|\nu_2>\, .\ee  However, because of the
ambiguities with the definition of flavor Fock creation-annihilation
operators discussed above these single particle states are
\emph{indirectly} related to the flavor \emph{fields}
$\nu_{e,\mu}(\vec{x},t)$ that enter in the standard model
Lagrangian. Furthermore, this \emph{single particle} definition does
not yield any information on a Fock representation of many particle
flavor states. A quantum statistical description of a neutrino gas
is intrinsically a \emph{many body} description, the total wave
function of an n-fermion system must be completely antisymmetric
under pairwise exchange. The second quantized Fock representation
allows a systematic treatment of the many particle aspects, in
particular in quantum statistical mechanics a \emph{distribution}
function is an expectation value of Fock number operator  in the
density matrix. Therefore simply \emph{defining} flavor states as in
eqn.(\ref{deflavor}) does not yield a complete information on the
many particle nature of a neutrino gas. A systematic study of the
many particle aspect of the dynamical evolution of a dense gas of
flavor neutrinos with a physically motivated definition of flavor
states even in the non-interacting theory was presented in
ref.\cite{hoboydense} wherein subtle but important effects
associated with the non-trivial Bogoliubov coefficients in the
dynamics were studied. Another subtlety emerges when a chemical
potential is assigned to ``flavor states'', a chemical potential is
a thermodynamic variable conjugate to a conserved particle number.
Even in the free theory, in absence of weak interactions, flavor
number is not conserved if neutrinos mix and as a result a chemical
potential for flavor neutrinos is not a well defined quantity even
for free mixed neutrinos. Dynamical aspects associated with this
issue were also studied in ref.\cite{hoboydense}.

 A counter argument to this critique would hinge on the fact that neutrino
masses are
 small on the relevant energy scales and mass differences are even smaller,
 therefore one can \emph{approximately} take all of the Dirac
 spinors to be \emph{practically} massless. Of course if all spinors
 $\mathcal{U},\mathcal{V}$ for mass and flavor eigenstates are taken
 to be massless the overlaps (\ref{overlapsUV}) yield $\mathcal{U}^\dagger
 \mathcal{U}=1; \mathcal{U}^\dagger \mathcal{V} =0$ and eqn. (\ref{deflavor})
becomes an
 identity. This point will become relevant below.

 While this approximation may be justified, it glosses over the main
 conceptual aspects and avoids the fundamental question of what
 precisely is a \emph{distribution  function} of flavor states. Such
 function includes information over all scales,   it
 yields the average occupation for \emph{all values of the momenta}, not
 just the high energy limit.

The main point of this discussion is that there are subtleties and
caveats in the kinetic description based on ``flavor states'' or
flavor matrix of densities, which involve Fock operators for flavor
states. While these subtleties and caveats may not invalidate the
broad aspects of the kinetic results, they cloud the interpretation
of the equilibrium state of neutrinos.

To highlight this point  consider an equilibrium situation in which
$n_{e\mu}(\vk)\equiv <\nu^\dagger_e(\vk)\nu_\mu(\vk)> =0$, for a
density matrix diagonal in the mass basis this means \be \cos\theta
\sin\theta\left[<\nu^\dagger_1(\vk)\nu_1(\vk)>-<\nu^\dagger_2(\vk)\nu_2(\vk)>\right]
=0 \, \Rightarrow <\nu^\dagger_1(\vk)\nu_1(\vk)> =
<\nu^\dagger_2(\vk)\nu_2(\vk)> \label{massdiag}\ee which in turn
leads to the result \be n_{ee}=<\nu^\dagger_e(\vk)\nu_e(\vk)> =
<\nu^\dagger_\mu(\vk)\nu_\mu(\vk)>=n_{\mu\mu} \ee

These conditions of ``flavor equalization'' are the \emph{same} as
those obtained in ref.\cite{barbieri} for the equilibrium solution
of the kinetic equations, although in that reference one active and
one sterile neutrino were studied. The condition (\ref{massdiag}) is
\emph{consistent}   with identical chemical potentials for the mass
eigenstates in the limit $m_1=m_2=0$. As discussed above, it is
precisely  taking $m_1=m_2=0$ that yields the correspondence between
the definition of the flavor states (\ref{deflavor}) and the
relation between the flavor and mass eigenstates \emph{fields} when
all spinors are taken to be massless. This is also the approximation
used in ref.\cite{rafsigl}  where flavor fields are expanded in the
basis of \emph{massless} spinors. These are precisely the
approximations invoked in the kinetic approach and correspond to
neglecting the neutrino masses. Restoring neutrino masses the off
diagonal correlation in the mass basis would be $n_{e\mu}(k) \propto
(m^2_1-m^2_2)/k^2$. Thus an interpretation of the kinetic results is
that the equilibrium state is described by a density matrix diagonal
in the mass basis with \emph{equal chemical potential} for the mass
eigenstates with an off diagonal correlation $n_{e\mu}\propto
(m^2_1-m^2_2)/k^2$ which is \emph{neglected} in the kinetic
approach.

Thus the equilibrium solution of the kinetic equation in
ref.\cite{barbieri}(see eqn. (12) in this reference) can be
interpreted as a \emph{confirmation} of the statement of
equilibration in the mass basis when the neutrino masses are
neglected, although in the ``flavor'' formulation of the kinetic
equations this information is not readily available.

The discussion in the previous section based on general aspects of
the full density matrix and a systematic perturbative expansion
avoids the caveats associated with the intrinsic ambiguities in the
definition of flavor states and suggests that equilibration leads to
a density matrix nearly diagonal in the \emph{mass basis}.

\vspace{1mm}

\textbf{Quantum Zeno effect:} References\cite{stod,fullerQZ,footQZ}
discuss the fascinating phenomenon of the Quantum Zeno effect or
``Turing's paradox''\cite{stod}. In the case of neutrino mixing,
this effect arises when the scattering rate is larger than the
oscillation rate. Since neutrinos are produced in weak interaction
vertices as ``flavor eigenstates'' when rapid collisions via the
weak interactions which are diagonal in the flavor basis prevent
oscillations, the states are effectively ``frozen'' in the flavor
basis\cite{stod}. This situation \emph{may} be expected at high
temperature. An order of magnitude estimate reveals that such a
possibility is \emph{not} available in the case under consideration,
with a large difference in the neutrino asymmetries. The argument is
the following: the oscillation frequency is given by \be  \Omega
\sim \frac{\delta m^2}{k}\,  \Bigg[\Big(\cos2\theta -
\frac{V(k)}{\delta
m^2}\Big)^2+\Big(\sin2\theta\Big)^2\Bigg]^{\frac{1}{2}} \;,\ee

where the matter potential \be V(k) \approx k \, G_F T^3 L \;,\ee
and $L$ is the neutrino asymmetry \emph{difference} between the two
generations of neutrinos. Our study relies on the possibility of
large asymmetries, namely $L \sim 1$. The decay rates  are of the
order \be\Gamma \approx G^2_F T^5 \;.\ee

The ``quantum zeno effect'' would operate provided $\Gamma >>
\Omega$. Even if  $V(k) >> \delta m^2$ which  can occur at high
temperatures, the oscillation frequency \be \Omega \sim G_F T^3 \,L
\ee and $\frac{\Gamma}{\Omega} \sim G_F T^2 /L <<1$  under the
assumptions invoked in this article, namely: i) $L\sim 1$, ii)
perturbation theory is valid in Fermi's effective field theory. A
reversal of this bound would entail a breakdown of the perturbative
expansion, and of the Fermi effective field theory. In perturbation
theory the bound is even stronger if $\delta m^2 >> V(k)$. Thus we
conclude that, under the conditions studied in this article,  the
quantum Zeno effect is not   effective in ``freezing'' the states in
the flavor basis and that oscillations and relaxation may indeed
result in a density matrix which is off diagonal in the flavor
basis.

\subsection{Main assumptions}\label{assumptions}

After the above discussion on the general aspects of equilibration
and the kinetic approach, we are in position to clearly state our
main assumptions. These are the following:

\begin{itemize}

\item{ {\bf i):} for $T
>> \,1\,\textrm{MeV}$ the electromagnetic and weak interaction rates
ensure that leptons are in    equilibrium in the early Universe,
namely their distribution functions are time independent. }

\item{{\bf  ii):}   the   results from the kinetic approach  in
refs.\cite{doleq}    indicate that for $T>>30\,\textrm{MeV}$
neutrino oscillations are suppressed and \emph{flavor} equilibration
via oscillations is \emph{not} operational. Therefore for $T>>
30\,\textrm{MeV}$ neutrinos are in equilibrium, and   there could be
large asymmetries in the neutrino sector consistent with the BBN and
CMB bounds in the \emph{absence} of oscillations.}

\item{{\bf iii):}   the   arguments presented above lead us to
   \emph{assume} that the   equilibrium state of the neutrino gas is described
by a
 density matrix which is nearly diagonal in the mass basis, and allow the
distribution functions of mass
 eigenstates to feature different chemical
potentials. As per the discussion above, this is consistent with the
interpretation of the equilibrium state \emph{after} equalization of
the chemical potentials discussed after eqn. (\ref{massdiag}).

Equation (\ref{propamix}) entails that the equilibrium off-diagonal
flavor propagator does not vanish.   The numerical study in
ref.\cite{doleq} shows that flavor equilibration with
$|\xi_\nu|\lesssim 0.07$ is established but for $T \sim
2\,\textrm{MeV}$, well below the scale of interest in our study,
clearly leaving open the possibility, which we assume here, of large
 asymmetries in the neutrino sector at a temperature much higher than that of
flavor equalization. In summary: the combination of the general
arguments suggesting that the equilibrium density matrix is nearly
diagonal in the mass basis, at least within the framework of
perturbation theory, along with the results of the kinetic approach
lead us to \emph{assume} that at high temperature $T \gtrsim
30\,\textrm{MeV}$ neutrinos are in equilibrium, the density matrix
is nearly diagonal in the mass basis and there could be large
asymmetries for the different mass states consistent with the bounds
in absence of oscillations. Equilibration of mass eigenstates
implies that the neutrino propagators in the mass basis only depend
on the time difference (translational invariance in time) and are
determined by the equilibrium distribution functions,  the
off-diagonal flavor propagator is given by eqn.(\ref{propamix}).
}

\end{itemize}

The \emph{dynamics} that leads to equilibration and the mechanism by
which substantial chemical potentials emerge is of course very
important and require a much deeper and detailed investigation as
well as an understanding of initial conditions for lepton
asymmetries. A consistent study should address the subtleties and
caveats associated with the notion of flavor states or Fock
operators. Such program is certainly beyond the scope and focus of
this article. Here we study the consequences of this assumption
  in a perturbative expansion in the mass basis up to one
loop order, namely up to  leading order in the weak and
electromagnetic interactions. In order to explore the main possible
consequences of neutrino equilibration with large asymmetries within
a simpler setting, we focus on the regime of temperature much larger
than the charged lepton masses.

We emphasize, that  our main observation, namely that charged
leptons mix if neutrinos mix is one of \emph{principle} and of a
general nature. Our purpose is to study the potentially novel broad
aspects of charged lepton mixing under these circumstances in the
simplest scenario, postponing a more detailed analysis to a future
study.

\section{Exploring the consequences}\label{mixing}

  In perturbation theory in the
electroweak interactions, the self-energy is computed in the basis
of mass eigenstates, hence it is off-diagonal in the flavor basis if
neutrinos are in equilibrium as mass eigenstates in the absence of
oscillations. Including the electromagnetic self-energy, and
introducing the charged lepton spinor fields $\nu_\alpha$ with
$\alpha\,,\,\beta=e,\mu$ the effective Dirac equation in the medium
for the space-time Fourier transforms of these fields is the
following\cite{hoboya} \be \left[\left(\gamma^0\,\omega-
\vec{\gamma}\cdot \vk\right)\delta_{\alpha\,\beta}
-\mathds{M}_{\alpha\,\beta}+ \Sigma^{em}_{\alpha\,\beta
}(\om,k)+\widetilde{\Sigma}^{NC}_{\alpha\,\beta
}(\om,k)+\Sigma^{NC}_{\alpha\,\beta
}(\om,k)\,L+\Sigma^{CC}_{\alpha\,\beta
}(\om,k)\,L\right]\,\psi_\beta (\omega,k)=0 , \label{EqMotFT} \ee
where $\mathds{M} = \textrm{diag}(M_e,M_\mu)$ is the charged lepton
mass matrix and $L=(1-\gamma^5)/2$.

\subsection{Electromagnetic Self-Energy}

 The leading electromagnetic
contribution to the charged lepton self-energy
$\Sigma^{em}_{\alpha\,\beta }(\om,k)$ for temperatures much larger
than the lepton masses is dominated by one-photon exchange in the
hard-thermal loop approximation\cite{htl,bellac}. As discussed
below, the temperature region of interest for substantial
charged-lepton mixing is $T \sim  \textrm{Gev}$, therefore we will
neglect corrections of order $M^2_e/T^2;M^2_\mu/T^2 \ll 10^{-2}$
(which already multiply one power of $\alpha$) to leading order.

Quark-lepton chemical equilibrium may lead to charged lepton
chemical potentials as large as those for   neutrinos, therefore we
allow for arbitrary charged lepton chemical potentials with the
possibility that $\mu_{e,\mu}/T \sim \mathcal{O}(1)$. The
self-energy $\Sigma^{em}_{\alpha\,\beta }(\om,k)$ is diagonal in
flavor space $\Sigma^{em} (\om,k)=\textrm{diag}\left(\Sigma^{em}_e
(\om,k),\Sigma^{em}_\mu (\om,k)\right)$. The matrix elements are
given by ($f=e,\mu$) \cite{htl,bellac} \be \Sigma^{em}_f(\om,k)=
-\gamma^0\,\left[\frac{m^2_f}{k}Q_0\left(\frac{\om}{k}\right)-i\Gamma\right]+\vec\gamma\cdot
\widehat{\bf{k}}\, \frac{m^2_f}{k}\left(\frac{\om}{k}\,
Q_0\left(\frac{\om}{k}\right)-1\right) \label{Sigem}~~;~~
\,Q_0(x)=\frac{1}{2}\ln\frac{x+1}{x-1}\ee with \be\label{Tmasses}
m^2_f=
\frac{\pi}{2}\,\alpha\,T^2\left[1+D(\xi_f)\right]~~;~~f=e,\mu. \ee
The function  \be D(\xi_f) = \frac{4}{\pi^2}\int^\infty_0 x dx
\left[\frac{1}{e^{x-\xi_f}+1}-\frac{1}{e^{x }+1} \right]~~;~~\xi_f =
\frac{\mu_f}{T} \ee is   monotonically increasing      with $-0.196
\leq D(\xi) \leq 0.4,~\textrm{for}~-1\leq \xi \leq 1$.

The leading order damping rate $\Gamma$ in $\Sigma^{em}$  emerges
from threshold infrared divergences associated with the emission and
absorption of soft, transverse magnetostatic
photons\cite{iancu,anomadamboya} and is insensitive to the masses
for  temperature larger than the mass of the charged leptons. In an
abelian plasma transverse photons do not acquire a magnetic mass and
are only screened \emph{dynamically} by Landau
damping\cite{iancu,anomadamboya,bellac}. Detailed work in high
temperature QED plasmas\cite{iancu,anomadamboya,bellac} reveals that
the exchange of soft magnetostatic photons yields an anomalous
damping of fermionic excitations, which can be accurately described
by the damping rate given by \be \Gamma = \alpha T \,
\ln\left(\frac{\omega_p}{\alpha T}\right)\label{Gamma}.\ee The
plasma frequency $\omega_p$ is determined by the photon polarization
fermion loop with all relativistic fermionic species: for $T\sim
\,\textrm{GeV}$ with electron, muon and three light quark degrees of
freedom we find to leading logarithimic order \be\label{Gammafin}
\Gamma= \alpha T \ln \left(\frac{8\pi}{3e}\right) \sim 3.3\, \alpha
T. \ee A collisional contribution to the charged lepton damping rate
$\Gamma$ is of order $\alpha^2 T$\cite{iancu} and will be neglected
to leading order in $\alpha$.

\subsection{Charged and neutral currents self-energy}

\begin{figure}
\begin{center}
\includegraphics[height=2in,width=2in,keepaspectratio=true]{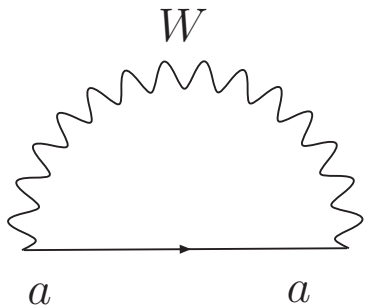}
\caption{Charged current self energy $\Sigma_a(\omega,k)$ with
$a=1,2$ corresponding to mass eigenstate neutrinos in the fermion
line. The external lines correspond to $e,\mu$ charged leptons with
frequency $\omega$ and momentum $k$.} \label{sigma}
\end{center}
\end{figure}

As depicted in fig. (\ref{loopmixing}) to lowest order in
perturbation theory in the interaction picture of $H_0$,  there is a
flavor off-diagonal contribution to the charged lepton self energy,
$\Sigma_{e,\mu}$ given by a W-boson exchange and an internal
off-diagonal fermion propagator $<
\overline{\nu}_e(\vec{x},t)\nu_{\mu}(\vec{x}',t') >$. In the mass
basis this  propagator is given by eqn. (\ref{propamix}) and the
general form of the corresponding  self energy contribution is
depicted in fig.\ref{sigma} where the internal fermion line
corresponds to a mass eigenstate neutrino $\nu_a$ in  equilibrium
with chemical potential $\mu_a = T\,\xi_a$.

We focus on the temperature regime $T \ll M_W$ and obtain the
charged current contribution to the self-energy up to leading order
in a local expansion in the frequency and momentum of the external
leptons neglecting terms proportional to  $m_a/M_W \lesssim
10^{-17}$. The general expressions for charged and neutral current
self-energies are given in references\cite{notzold,hoboya} where we
refer the reader for more details.

 Denoting by
$\Sigma^{CC}_a(\omega,k)$ the one-loop charged current self-energy
with internal neutrino line corresponding to  a mass eigenstate
$\nu_a$ displayed in fig.\ref{sigma},  the matrix
$\Sigma^{CC}_{\alpha,\beta}(\omega,k)$ in eqn. (\ref{EqMotFT}) has
the following entries \bea\label{sigmas}\Sigma^{CC}_{e,e} & = &
C^2\,\Sigma_1 +
S^2\,\Sigma_2 \nonumber \\
\Sigma^{CC}_{\mu,\mu} & = & S^2\,\Sigma_1 + C^2\,\Sigma_2 \nonumber \\
\Sigma^{CC}_{e,\mu} & = & \Sigma^{CC}_{\mu,e} = -CS
\left(\Sigma_{1}-\Sigma_{2}\right)\eea  consistently with a
perturbative calculation in the mass basis to lowest order in the
weak interactions $C=\cos \theta \,;\,S=\sin \theta $ and $\theta$
is the neutrino vacuum mixing angle. A fit to the solar and
KamLAND\cite{kamland} data yields $\tan^2\theta \approx 0.40$.

The off diagonal element $\Sigma^{CC}_{e,\mu}$ in eqn.
(\ref{sigmas}) is responsible for charged lepton mixing. The flavor
off-diagonal propagator in this self-energy is precisely given by
eqn. (\ref{propamix}). The form of this off-diagonal self-energy
makes clear that there is mixing provided $\theta \neq 0,\pi/2$
\emph{and} $\Sigma_1-\Sigma_2\neq 0$. The calculation of the
self-energy $\Sigma^{CC}_{e,\mu}$ in the \emph{vacuum} is standard:
it is performed in the interaction picture of the true basis of
in-out states, these are \emph{mass eigenstates}. In this case the
difference of the self-energies is determined solely by the neutrino
mass difference, therefore in the vacuum $\Sigma^{CC}_{e,\mu}
\propto G_F \Delta m^2 \sim 10^{-27}$ and the mixing between charged
leptons is negligible.

 The main point of our study is that in the
medium in equilibrium  with large neutrino asymmetries for the mass
eigenstates, charged lepton mixing may be substantial. The
propagating modes in the medium are determined by the poles of the
exact propagator. An off diagonal self-energy $\Sigma_{e\mu}$
entails that the charged lepton propagating modes are admixtures of
electron and muon degrees of freedom.  We now study this possibility
in detail when the temperature is much larger than the lepton
masses. We focus on this case for simplicity in order to extract the
main features of the phenomenon and to highlight the main steps in
the calculation.

We are only interested in the real part of the self-energies since
at this order the imaginary part vanishes on the charged lepton mass
shells. From the results obtained in\cite{hoboya}, for any loop with
a lepton with mass $m $ in the high temperature limit $T\gg m $ we
obtain \be \textit{Re}\,{\Sigma} (\omega,k)=\gamma^0 \; \sigma^0
(\omega, k)-\vec\gamma\cdot \widehat{\bf{k}} \; \sigma^1
 (\omega,k).\label{ReSelf} \ee   For $M_{W,Z}\gg T,\omega,k$    we
find \bea {\sigma}^0 (\omega,k)&=& -\frac{3 \;
g\,G_F\,n_\ga}{\sqrt2}L + \frac{7 \; \pi^2}{15 \; \sqrt2}\frac{g
\,G_F\;
\om \; T^4}{M_B^2}\,I  \; , \\
{\sigma}^1 (\omega,k)&=& -\frac{7 \; \pi^2}{45 \; \sqrt2} \; \frac{g
\,G_F\;  k \; T^4}{M^2_B}\,I  \, , \label{sigmaW} \eea where $g$ is
the appropriate factor for charged or neutral currents, $L$ is the
asymmetry for the corresponding lepton and $M_B=M_{W,Z}$ for charged
or neutral currents,   $ n_\ga= {2} \; \zeta(3) \; T^3 /{\pi^2} $,
and \be I = \frac{120}{7\pi^4} \int_0^\infty \frac{x^3}{e^{x-\xi
}+1}\, dx \,.\ee The neutrino asymmetries are given by $L_a =
(\pi^2/12\zeta(3))(\xi_a+\xi^3_a/\pi^2)$. In the absence of
oscillations the combined analysis from CMB and BBN yield an upper
bound on the asymmetry parameters $|\xi_{e}|\lesssim 0.1,|\xi_\mu|
\sim 1$ for \emph{flavor neutrinos}\cite{hansen,kneller} which we
\emph{assume} to imply a similar bound on the asymmetries for the
mass eigenstates, $\xi_a$. The validity of this  assumption in free
field theory is confirmed by the analysis in ref.\cite{hoboydense}.
From (\ref{sigmas}) and the above results the following general form
for the charged and neutral current self-energies is obtained

\be \label{ReSig} \textit{Re}\,\Sigma(\om,k) =\frac{3\;
G_F\,n_\ga}{2\sqrt2}\,[\gamma^0 \mathds{A}(\om )-\vec{\ga}\cdot\uvk
\,\mathds{B}( k)] \ee \noindent where $\mathds{A}(\om,k)$ and
$\mathds{B}(\om, k)$ are $2\times2$ matrices in the charged lepton
flavor basis given by \bea \mathds{A}(\om )= \left(
                   \begin{array}{cc}
                    A_{ee}(\om ) & A_{e\mu}(\om) \\
                    A_{e\mu}(\om)& A_{\mu\mu}(\om ) \\
                   \end{array}
                 \right)~;~~
\mathds{B}(k) = \left(
                   \begin{array}{cc}
                    B_{ee}( k) & B_{e\mu}(k) \\
                    B_{e\mu}(k)& B_{\mu\mu}( k) \\
                   \end{array}
                 \right). \eea

 \noindent where the matrix elements are given by

\bea A^{CC}_{ee}(\om) &=& -\Bigg[\,L_+ + \cos2\theta \,L_- - \frac{7
\;
\pi^4}{90 \; \zeta(3)}\frac{\om \; T}{M_W^2}\, (I_+\,+ \cos2\theta\,I_-)\Bigg]\\
 A^{CC}_{\mu\mu}(\om) &=& -\Bigg[\,L_+ - \cos2\theta \,L_- - \frac{7 \;
\pi^4}{90 \; \zeta(3)}\frac{\om \; T}{M_W^2}\,(I_+\,- \cos2\theta\,I_-)\Bigg]\\
A^{CC}_{e\mu}(\om) &=&\,  \sin2\theta \;\Bigg[ L_- - \frac{7 \;
\pi^4}{90 \; \zeta(3)}\frac{\om \; T}{M_W^2}\, I_-\Bigg]\eea  and
\bea B^{CC}_{ee}(k) &=& -\frac{7\pi^4}{270\,\zeta(3)}\,
\frac{k \; T}{M^2_W} (I_+\,+ \cos2\theta\,I_-) \\
 B^{CC}_{\mu\mu}(k) &=& -\frac{7\pi^4}{270\,\zeta(3)}\,
\frac{k \; T}{M^2_W} (I_+\,- \cos2\theta\,I_-) \\
B^{CC}_{e\mu}(k)&=&  \frac{7\pi^4}{270\,\zeta(3)}\, \frac{k \;
T}{M^2_W} \sin2\theta\,I_-\, , \eea  where we have introduced
$L_\pm= L_1\pm L_2, I_\pm= I_1\pm I_2$. The neutral current
contributions are flavor diagonal and given by \bea A^{NC}_{ee}(\om)
&=& -(1-4\,\sin^2\theta_w)\,\Bigg[\,\mathcal{L}_e   - \frac{7 \;
\pi^4}{90 \; \zeta(3)}\frac{\om \; T}{M_W^2}\,
\cos^2\theta_w\,I_e\Bigg]-\frac{4}{3} \,\sum_f g^v_f\,L_f \\
 A^{NC}_{\mu\mu}(\om) &=&
-(1-4\,\sin^2\theta_w)\,\Bigg[\,\mathcal{L}_\mu   - \frac{7 \;
\pi^4}{90 \; \zeta(3)}\frac{\om \; T}{M_W^2}\,
\cos^2\theta_w\,I_\mu\Bigg]-\frac{4}{3} \,\sum_f g^v_f\,L_f\\
\widetilde{A}^{NC}_{ee}(\om) &=&
-4\,\sin^4\theta_w\,\Bigg[\,\mathcal{L}_e - \frac{7 \; \pi^4}{90 \;
\zeta(3)}\frac{\om \; T}{M_W^2}\,
\cos^2\theta_w\,I_e\Bigg]+\frac{8}{3}\,\sin^2\theta_w \,\sum_f
g^v_f\,L_f \\ \widetilde{A}^{NC}_{\mu\mu}(\om)   & = &
-4\,\sin^4\theta_w\,\Bigg[\,\mathcal{L}_\mu - \frac{7 \; \pi^4}{90
\; \zeta(3)}\frac{\om \; T}{M_W^2}\,
\cos^2\theta_w\,I_\mu\Bigg]+\frac{8}{3}\,\sin^2\theta_w \,\sum_f
g^v_f\,L_f \,,\eea  and \bea B^{NC}_{ee}(k) &=&
-\frac{7\pi^4}{270\,\zeta(3)}\,
\frac{k \; T}{M^2_W}(1-4\,\sin^2\theta_w)\,\cos^2\theta_w \,I_e \\
 B^{CC}_{\mu\mu}(k) &=& -\frac{7\pi^4}{270\,\zeta(3)}\,
\frac{k \; T}{M^2_W}(1-4\,\sin^2\theta_w)\,\cos^2\theta_w \,I_\mu \\
\widetilde{B}^{NC}_{ee}(k) &=&
\frac{4\,\sin^4\theta_w~B^{NC}_{ee}(k)}{(1-4\,\sin^2\theta_w)} ~~;~~
\widetilde{B}^{NC}_{\mu\mu}(k)  =
\frac{4\,\sin^4\theta_w~B^{NC}_{\mu\mu}(k)}{(1-4\,\sin^2\theta_w)}
\,, \eea where $g^v_f \;,\;L_f$ are the vector coupling and
asymmetry of fermion species $f$, $\theta_w$ is the Weinberg angle
and $\mathcal{L}_{e,\mu}$ are the \emph{charged} lepton asymmetries.
The non-vanishing off-diagonal matrix elements $A^{CC}_{e
\mu}\,,\,B^{CC}_{e\mu}$ lead  to charged lepton mixing and
oscillations. It is convenient to combine the charged leptons into a
doublet of Dirac fields \be \psi(\om,k)=\left(
                  \begin{array}{c}
                    \psi_e(\om,k) \\
                    \psi_{\mu}(\om,k) \\
                  \end{array}
                \right) \; .\ee
 In the chiral representation the left and right handed
 components of the Dirac doublet are written as linear combinations
 of  Weyl spinors $v^{(h)}$ eigenstates of the helicity operator
$\vec{\sigma}\cdot \uvk$ with eigenvalues $h=\pm 1$, as
follows\cite{hoboya}
  \be \psi_L = \sum_{h=\pm1} \left(
                        \begin{array}{ c}
                            0 \\
                          v^{(h)}\otimes \varphi^{(h)} \\
                        \end{array}
                      \right) ~~;~~ \nu_R = \sum_{h=\pm1} \left(
                        \begin{array}{ c}
                            v^{(h)}\otimes \xi^{(h)} \\
                          0 \\
                        \end{array}
                      \right) \ee
\noindent The left handed doublet \be \varphi^{(h)} (\omega,k) =
\left(
         \begin{array}{c}
           \l^{(h)}_{e } (\omega,k)\\
           \l^{(h)}_{\mu  }(\omega,k) \\
         \end{array}
       \right) \label{flavdou} \; , \ee obeys the following
       effective Dirac equation in the medium to leading order in
$\alpha,G_F$\cite{hoboya} \be \label{eignEqL}
\Bigg\{\left[(\om+i\Gamma)^2-k^2 \right]\mathds{1}+ \frac{3\;
G_F\,n_\ga}{2\sqrt2}\Bigg(2\omega\,\widetilde{\mathds{A}}^{NC}-
2k\,\widetilde{\mathds{B}}^{NC} +(\om-h k)(\mathds{A}+h
\mathds{B})\Bigg)- \widetilde{\mathds{M}}^2 \Bigg\}\varphi^{(h)}
(\omega,k)=0 \; , \ee where $ \widetilde{\mathds{M}}^2 =
\textrm{diag}\left(M^2_e+2m^2_e , M^2_\mu+2m^2_\mu    \right) $
where $m^2_{e,\mu}$ are given by eqn. (\ref{Tmasses}) and to avoid
cluttering of notation  $\mathds{A}\,,\, \mathds{B}$ are the
\emph{sums} of the charged and neutral current contributions. To
leading order the right handed doublet is determined by the
relation\cite{hoboya} \be \label{xieq} \xi^{(h)}(\omega,k) =
-\mathds{M} \; \frac{(\om +h \; k) }{\omega^2-k^2} \;
\varphi^{(h)}(\omega,k) \; . \ee The propagating modes in the medium
are found by diagonalization of the above Dirac equation. Let us
introduce a doublet of \emph{collective modes} in the medium \be
\label{massei} \chi^{(h)}(\om,k) = \left(
                  \begin{array}{c}
                    \l^{(h)}_1(\om,k) \\
                    \l^{(h)}_2(\om,k) \\
                  \end{array}
                \right) \; ,
\ee \noindent related to the flavor doublet
$\varphi^{(h)}(\omega,k)$ by a unitary transformation $ U^{(h)}_m $
with

\be \label{mediumrotmat} U^{(h)}_m = \left(
          \begin{array}{cc}
            \cos\theta_m^{(h)} & \sin \theta_m^{(h)} \\
            -\sin \theta_m^{(h)} & \cos \theta_m^{(h)} \\
          \end{array}
        \right) \; , \ee

\be  \label{unitrafo2} \varphi^{(h)}(\omega,k) = U^{(h)}_m \;
\chi^{(h)}(\omega,k) . \ee and a similar transformation for the
right handed doublet $\xi^{(h)}(\omega,k)$,  where the mixing angle
$\theta^{(h)}_m$ depend  on $h,k$ and $\omega$. The eigenvalue
equation in diagonal form is given by \be  \Bigg\{
(\omega+i\Gamma)^2-k^2 +\frac12 \; S_h(\omega,k)-\frac12 \;
(M^2_e+M^2_\mu+2\,m^2_e+2\,m^2_\mu)+\frac12 \; \Omega_h(\om,k)\left(
          \begin{array}{cc}
           1 & 0 \\
            0 & -1 \\
          \end{array}
        \right)\Bigg\}\chi^{(h)}(\omega,k)
          =0 \label{psimass}  \; ,
\ee

\noindent  where $S_h(\omega,k)$, $\Delta_h(\om,k) $ and
$\Omega_h(\om,k)$ are respectively given by

\bea S_h(\omega,k) & = & \frac{3\; G_F\,n_\ga}{2\sqrt2}\Bigg\{(\om-h
k)\left[A_{\mu\mu}(\omega )+A_{ee}(\omega ) + h( \; B_{ee}(
k)+B_{\mu\mu}( k))\right]+\nonumber\\ && 2\omega
(\widetilde{A}^{NC}_{\mu\mu}+
\widetilde{A}^{NC}_{ee})-2k(\widetilde{B}^{NC}_{\mu\mu}
+ \widetilde{B}^{NC}_{ee})\Bigg\}\label{Sofome} \; , \\
\Delta_h(\omega,k) &=&\frac{3\;
G_F\,n_\ga}{2\sqrt2}\Bigg\{(\om-hk)[\, A_{\mu\mu}(\omega
)-A_{ee}(\omega )\,+h(B_{\mu\mu}(\omega )-B_{ee}(\omega
))]+\nonumber\\ && 2\omega (\widetilde{A}^{NC}_{\mu\mu}-
\widetilde{A}^{NC}_{ee})-2k(\widetilde{B}^{NC}_{\mu\mu}
- \widetilde{B}^{NC}_{ee})\Bigg\} \label{Deltah} \; , \\
\Omega_h(\om,k)&=&\Bigg(\left[\,\delta \widetilde{M}^2
-\Delta_h(\om,k)\,\right]^2+\left[\,2(\om-hk)(A_{e\mu}+h\,B_{e\mu})\,\right]^2\Bigg)^{\frac12}.
\eea where $\delta \widetilde{M}^2=M^2_\mu-M^2_e+2m^2_\mu-2m^2_e$.
The mixing angle in the medium is determined by the relations \be
\label{sincosmix}\sin2\theta^{(h)}_m =
-\frac{2(\om-hk)\,(A_{e\mu}+h\,B_{e\mu})}{\Omega_h(\om,k)}~~;~~\cos2\theta^{(h)}_m
=  \frac{\delta \widetilde{M}^2 -\Delta_h(\om,k)}{\Omega_h(\om,k)}.
 \ee where $\omega$ must be replaced by the solution of the
 eigenvalue equation (\ref{psimass}) for each collective mode. A
 remarkable convergence of scales emerges for $T \sim 5  \,
 \textrm{GeV}$: \emph{if} the neutrino asymmetry $|L_-| \sim 1$ then
 for nearly thermalized relativistic charged leptons with $\omega \sim -h k$
with $k \sim
 T$, all of the terms in the expression for $\Omega_h(\om,k)$
 are of the same order, namely,  $|\Delta_h(k,k)| \sim |T A_{e\mu}| \sim
\delta \widetilde{M}^2  $. For relativistic
 leptons $|\omega|$ can be replaced by $k$ in the arguments of the
 functions $A,B$ to leading order in $\alpha,G_F$.

 For $M_{\mu}\ll \omega,k, T \ll M_W$ we find that the leading contribution to
 $\Delta_h$ is given by
\be \Delta_h(\omega,k)  \simeq 1.2\,10^{-5}\,
\Bigg(\frac{T}{\mathrm{GeV}}\Bigg)^4 \,\Bigg[\Bigg(\frac{\omega -
h\,k}{2T}\Bigg) L_-\, \cos2\theta - 0.29\,
\Bigg(\frac{6.3\omega-h\,k}{7.3 T}
\Bigg)\,\Bigg(\mathcal{L}_\mu-\mathcal{L}_e \Bigg) \Bigg]
\;(\mathrm{GeV}^2)\,, \ee where we have used the value
$\sin^2\theta_w =0.23$. A resonance in the mixing angle occurs for
$\Delta_h = \delta \widetilde{M}^2$. The typical momentum of a
lepton in the plasma is $k \sim T$, therefore in the temperature
regime $T \ll M_W$ wherein our calculation is reliable, a resonance
is available $\omega_{e,\mu}(k) \sim -h\, k \sim -h\,T$ \emph{if}
the neutrino asymmetry is close to the upper bound. Taking the
values for $|\xi_a| $ inferred from the upper bounds from combined
CMB and BBN
 data in absence of oscillations\cite{hansen,kneller}  and the fit $\tan^2\theta
\sim 0.40$ from the combined solar and KamLAND data\cite{kamland}
 suggest the upper bound $|L_-\cos 2\theta| \sim 1$. With the asymmetry
parameters for the
 charged leptons $|\xi_f|$ smaller than or of the same order of $ |\xi_a|$,
resonant mixing may occur in the temperature range  $T \sim
 5
 \,\mathrm{GeV}$. Even when the asymmetries from charged leptons do
 not allow for a resonance or at lower temperature, it is clear that at high
temperature
  and for large neutrino asymmetries such that $L_- \sim 1$ there is a
\emph{large mixing angle} because of the
 convergence of scales. Hence at high temperature and large \emph{differences}
in the chemical potential
 for mass eigenstates, the propagating charged
 lepton collective excitations in the medium will be large
 admixtures of $e\,;\mu$ states. Consider a slightly off-equilibrium
 disturbance in the medium corresponding to an initial state describing an
inhomogeneous wave
 packet of electrons.
 The real time evolution of this state in the medium has to be
 studied as an initial value problem. Following the real time analysis presented
in ref.\cite{hoboya}, we find that if an initial state describes a
wave-packet of left handed  electrons of helicity $h$,  with
amplitude $l^h_e(0;k);k\gg M_{\mu}$ but no muons, the persistence
and transition probabilities are given by \bea \label{transproba}
P_{e\rightarrow e}(t;k) & = & |l^h_e(0;k)|^2\,e^{-2\Gamma t}
\Bigg[1-\sin^2(2\theta^h_m(k))\sin^2\left[ \frac{\Omega
(k)}{4k}\,t\right] \Bigg]
\\P_{e\rightarrow \mu}(t;k) & = &
|l^h_e(0;k)|^2\,e^{-2\Gamma t}
 \sin^2(2\theta^h_m(k))\sin^2\left[ \frac{\Omega (k)}{4k}\,t\right]
~~;~~\Omega(k)=\Omega_h(k,k)\,.\eea The exponential prefactor
reveals the \emph{equilibration} of the charged lepton distribution
with the equilibration rate $2\Gamma$\cite{bellac,anomadamboya}. It
is also remarkable that $\Gamma \sim \Omega(k)/k$ in the temperature
and energy regime of relevance for the resonance $k\sim T \sim 5
\,\textrm{GeV}$. Therefore we conclude that during the equilibration
time scale of charged leptons, there is a substantial transition
probability. Collisional contributions are of order $\alpha^2 T$ or
$G^2_F T^5$ for electromagnetic or weak interaction processes
leading to collisional relaxation time scales far larger than the
oscillation scale for $T \sim 5\,\textrm{GeV}$.

In the radiation dominated phase for $M_W \gg T$ as discussed here,
we
 find that the ratio of the oscillation to the expansion time
 scale $H\tau_{osc} \lesssim 10^{-16} (T/\mathrm{GeV})^3 \ll 1$,
 namely oscillation of mixed charged leptons occur on time scales
 much shorter than the expansion scale and the cosmological expansion  can be
considered adiabatic.

 \subsection{Remarks: beyond perturbation theory}\label{PT}

 We have focused on the high temperature limit to provide a detailed
 calculation within a simpler scenario, to extract the main aspects
 of the phenomenon and to highlight the main steps of the
 calculation. However, it is clear that a similar calculation can be
 performed at much lower temperature and the point of principle is
 still valid under the assumption of an equilibrium density matrix diagonal in
the mass basis: there \emph{could} be substantial charged lepton
 mixing if there are large chemical potential differences between
 the distribution functions of mass eigenstates. As per the
 discussion above, this is \emph{not} the only scenario that yields
 substantial charged lepton mixing, the general condition is that
 the off diagonal self energy $\Sigma_{e\mu}$
 in eqn. (\ref{ave}) be non-zero (and large). The off diagonal
 expectation value $< \nu_{eL}\overline{\nu}_{\mu L} >$ must in
 general be found from the equilibrium solution of a kinetic
 equation, but with a consistent treatment that avoids the caveats
 and subtleties discussed in section (\ref{kinetics}).

 The arguments in favor of an equilibrium density matrix diagonal
 (or  nearly so) in the mass basis, and the specific calculation
 described above relied on a perturbative expansion in the
 interaction picture of the unperturbed Hamiltonian $H_0$ which
 includes the neutrino mass matrix. There are possible caveats in
 the validity of perturbation theory, particularly in the case where
 medium effects lead to large corrections to the \emph{single
 particle} states. For example, a large ``index of refraction''
 arising from forward scattering with particles in the medium may
 lead to non-perturbative changes in the properties of the single
 particle basis. The lowest order contribution to the self-energy
 from forward scattering has been obtained in ref.\cite{notzold},
 these are in general dependent on the energy of the neutrinos.
 Including these corrections in a perturbative approach entails
 summing the geometric Dyson series for the one-particle irreducible
 self energy in the neutrino propagators. This case is akin to the
 generation of a thermal mass from forward scattering in a scalar
 $\phi^4$ field theory at finite temperature\cite{bellac}, when this
 thermal mass is larger than the zero temperature mass there is a
 large modification in the propagating single particle modes in the
 medium. In the scalar field theory case   a self-consistent
 re-arrangement of the perturbative expansion consists in adding the
 thermal mass to the \emph{unperturbed} Hamiltonian and at the same
 time a counterterm in the interaction part. The free single particle
propagators now include
 the thermal mass term, and in order to avoid double counting, the
 counterterm in the interaction Hamiltonian cancels the
 contributions that yield the thermal mass corrections
 systematically order by order in the perturbative expansion. We
 propose a similar strategy to include the medium modifications to
 the propagating single particle modes in the medium. In
 references\cite{notzold,hoboya} it is found that the forward
 scattering contributions to the effective Hamiltonian in the medium
 are of the form

 \be \delta H = \gamma^0 \mathds{A}(k)-\vec{\gamma}\cdot
 \hat{\vec{k}}\,\mathds{B}(k) \ee with $\mathds{A}(k);\mathds{B}(k)$
 momentum dependent  matrices in flavor space,   their explicit
 expressions are given in refs.\cite{notzold,hoboya}. A
 re-arrangement of the perturbative expansion results by writing \be
 H= \widetilde{H}_0 + \widetilde{H}_{int} \ee where $\widetilde{H}_0 =
 H_0+\delta H$ and $\widetilde{H}_{int}=H_{int}+H_c$ where the
 ``counterterm'' Hamiltonian $H_c=-\delta H$ systematically cancels the
 forward  scattering corrections to the self-energies consistently
 in the perturbative expansion. The new ``free'' Hamiltonian
 $\widetilde{H}_0$ includes self-consistently the modifications to
 the propagating single particle states from the in-medium index of
 refraction. The field operators are now written in the basis of the
 solutions of the Dirac equation from the new Hamiltonian and
 finally the interaction is written in terms of these fields. Thus
 the perturbative expansion is re-organized in terms of the single
 particle propagating modes in the medium. The main complication in
 this program is that the mixing angles in the medium which
 determine the single particle propagating modes, are energy
 dependent, this introduces a non-locality in the interaction
 vertices which now become dependent on the energy and momentum
 flowing into the vertex.

 If there is substantial charged lepton mixing, such phenomenon in
 turn affects the index of refraction for neutrinos in the medium
 and possibly the equilibration dynamics of neutrinos. A
 self-consistent treatment of the charged lepton mixing and neutrino
 mixing and relaxation would be required to understand the dynamical
 aspects of neutrino and charge lepton equilibration. This task is
 beyond the goals and focus of this article.

 \section{Conclusions and discussions}\label{conclu}

  In this article we focused on studying the possibility of charged
 lepton mixing as a consequence of neutrino mixing at high
 temperature and density in the early Universe.  There are  three main points in
this article:

\begin{itemize}

 \item{ (I)    We establish that a general
 criterion  for charged lepton mixing as a consequence of neutrino
 mixing is that there must be off diagonal correlation of flavor fields in the
  density matrix. We identified \emph{one}   possible case in
 which there could be   charged lepton mixing: that the
 equilibrium density matrix be nearly diagonal in the mass basis.
 This is the case in the vacuum, but in this case the smallness of
 the neutrino masses entails that charged lepton mixing is
 negligible. We argued that this effect can be enhanced in a medium
 if the density matrix is nearly diagonal in the mass basis
 with large and different chemical potential for mass eigenstates.
  While this is not the only case in which
 charged lepton mixing can occur, it is one in which we can provide a definite
calculation
 to assess charged lepton mixing. }

 \item{ (II)    We have given general arguments to suggest that within the realm
of validity of
 perturbation theory, the equilibrium density matrix must be nearly diagonal in
the mass basis.
 We have critically re-examined the kinetic approach to neutrino mixing and
relaxation in a medium
 at high temperature and density and highlighted several caveats and subtleties
with flavor states, and or Fock operators
 associated with these states that cloud the interpretation of the density
matrix.
 We   argue  that an equilibrium solution of the kinetic equations describing
``flavor
 equalization''\cite{barbieri}
 can be interpreted as a confirmation that the density
 matrix is nearly diagonal in the mass basis. This interpretation leads us to
the main and only assumption, namely
 that \emph{before} ``flavor equalization'' for $T  \gtrsim 30 \,\textrm{MeV}$
neutrinos are in equilibrium, the density matrix is nearly diagonal
 in the mass basis but with distribution functions for mass eigenstates with
large and different chemical potentials in
 agreement with the bounds from BBN and CMB in absence of oscillations.   }

 \item{ (III)  Under this assumption and the validity of perturbation theory we
have provided a
 definite calculation of charged lepton mixing. While the general criterion for
charged lepton mixing does not imply that
 this is the \emph{only} case in which charged leptons mix, it is a scenario
that allows a definite calculation to assess the
 phenomenon in a quantitative manner.    }

 \end{itemize}

 In conclusion, under the assumption that the mass eigenstates of  mixed
neutrinos in the early Universe
 are in thermal equilibrium    with different chemical potentials for
$T>>30\,\textrm{MeV}$, \emph{before} oscillations
 establish  the equalization of flavor asymmetries\cite{doleq}    neutrino
 mixing leads to charged  leptons mixing with large mixing angles in the
 plasma. We explored this possibility by obtaining the leading order
contributions to the charged lepton self
 energies in the high temperature limit.  If the upper bounds on
 the neutrino asymmetry parameters from BBN and CMB without oscillations is
assumed along
 with the fit for the vacuum mixing angle for two generations from
 the KamLAND data, we find that charged leptons mix resonantly in
 the  temperature range $T \sim  5 \,\mathrm{GeV}$ in the early
 Universe. The electromagnetic damping rate is of the same order
 as the oscillation frequency in the energy and temperature regime
 relevant for the resonance suggesting a substantial transition
 probability during equilibration.  The cosmological
 expansion scale is much larger than the time scale of charged lepton
 oscillations. Although we assumed the validity of perturbation
 theory we recognized possible caveats in the high temperature
 limit arising from  potentially large corrections to the single-particle
propagating
 modes from the in-medium index of refraction. We proposed a
  re-organization of the perturbative expansion that includes the
 correct single-particle propagators self-consistently. We
 have focused on the high temperature limit as a simpler scenario to
 assess charged lepton mixing, however, the calculation can be performed at
 lower temperatures with the corresponding technical
 complications associated with the lepton masses. While at much
 lower temperatures there is no resonant mixing of charged leptons,
 the results of the calculation establish a point of principle,
 namely that  for large chemical potential differences in the distribution
 function of mass eigenstates of neutrinos, the charged lepton
 propagating modes in the medium will be admixtures of the electron
 and muon degrees of freedom with non-vanishing mixing angle.

 We believe that the phenomenon of charged lepton mixing in a medium
 warrants a deeper and thorough investigation. Our study also raises
relevant questions on the kinetic approach: a consistent description
of the kinetics of neutrino oscillation and relaxation avoiding the
caveats associated with flavor states and or Fock operators
associated with these states. Furthermore, substantial charged
lepton mixing also suggests that a dynamical description should
include  self-consistently both  neutrino and charged lepton mixing
in a full non-equilibrium treatment.
  These aspects as well as possible consequences of  charged lepton mixing for
leptogenesis   will be explored elsewhere.

\begin{acknowledgments} D.B. thanks the US NSF for  support under
grant PHY-0242134. C.\, M. Ho  acknowledges   support through the
Andrew Mellon Foundation. The authors thank G. Steigman and L.
Wolfenstein for an illuminating conversation.
\end{acknowledgments}


\begin{thebibliography}{99}

\bibitem{raffelt} G. G. Raffelt, \textit{Stars as Laboratories for
Fundamental Physics}, (The University of Chicago Press, Chicago,
1996); New Astron.Rev. \textit{46}, 699 (2002); J. N.
Bahcall,\textit{Neutrino Astrophysics},
 (Cambridge University Press, Cambridge, 1989).


\bibitem{panta} T. K. Kuo and J. Pantaleone, Rev. of Mod. Phys.
\textbf{61}, 937 (1989).

\bibitem{giunti}  S.M. Bilenky, C.
Giunti, J.A. Grifols, E. Masso, Phys. Rept. \textbf{379}, 69 (2003);
C. Giunti, Found. Phys. Lett. \textbf{17}, 103 (2004); S. M. Bilenky
and C. Giunti, Int. J. Mod. Phys. \textbf{A16}, 3931 (2001); S. M.
Bilenky, hep-ph/0402153 .

\bibitem{smirnov} A. Yu. Smirnov, Int.J.Mod.Phys. \textbf{A19}1180 (2004);
hep-ph/0306075; hep-ph/0305106.



\bibitem{haxton}  W. C. Haxton, in \textit{Boulder 1998, Neutrinos in physics
and astrophysics}, 432 (Ed. Paul Langacker,
 Singapore, World Scientific, 2000) (nucl-th/9901076), Wick C. Haxton, Barry R.
Holstein,  Am. J. Phys. \textbf{68}, 15 (2000).

\bibitem{grimus} W. Grimus, Lect. Notes in Phys. \textbf{629}, 169
(2004).

\bibitem{gouvea} A. de Gouvea, hep-ph/0411274 (TASI lectures on
neutrino physics).


\bibitem{fuller} M. J. Savage, R. A. Malaney and G. M. Fuller,
Astroph. Jour. \textbf{368}, 1, 1991; C. Y. Cardall and  G. M.
Fuller, astro-ph/9702001.


\bibitem{dolgov} A. D. Dolgov, Surveys High Energ.Phys. {\bf 17}, 91,
(2002); Phys. Rept. \textbf{370}, 333 (2002); Nuovo Cim. \textbf{117
B}, 1081, (2003); A.D. Dolgov, F.L. Villante, Nucl.Phys.
\textbf{B679} 261, (2004).

\bibitem{kirilova}  D. Kirilova and M. Chizhov, Nucl.Phys.Proc.Suppl.
\textit{100}, 360 (2001); Phys.Lett. \textbf{B393}, 375 (1997).

\bibitem{SN} G. M. Fuller, W. C. Haxton, G. C. McLaughlin
Phys.Rev.\textbf{D59}, 085005 (1999);   J. Pantaleone,  Phys.Lett.
\textbf{B342}, 250 (1995);  B. Jegerlehner, F. Neubig, G. Raffelt,
Phys.Rev.\textbf{D54}, 1194 (1996).

\bibitem{fukugita} M. Fukugita and T. Yanagida, Phys. Lett.
\textbf{B174}, 45 (1986); W. Buchmuller, P. Di Bari and M.
Plumacher, New J. Phys. \textbf{6},105 (2004); Annals Phys.
\textbf{315 }, 305 (2005); Nucl. Phys. \textbf{B665}, 445 (2003); W.
Buchmuller, R. D. Peccei and T. Yanagida, hep-ph/0502169.



\bibitem{notzold} D. Notzold and G. Raffelt, Nucl. Phys.
\textbf{B307}, 924 (1988).


\bibitem{dolivoDR} J. C. D' Olivo and J. F. Nieves, Int. Jour. of
Mod. Phys. \textbf{A11}, 141 (1996); J. C. D'Olivo, J. F. Nieves and
M. Torres, Phys. Rev. \textbf{D46}, 1172 (1992);  E. S. Tututi, M.
Torres and J. C. D'Olivo, Phys. Rev. \textbf{D66}, 043001 (2002).


\bibitem{barbieri} R. Barbieri and A. Dolgov, Nucl. Phys. \textbf{B
349}, 743 (1991).

\bibitem{enqvist} K. Enqvist, K. Kainulainen and J. Maalampi,
Nucl. Phys. \textbf{B349}, 754, (1991).

\bibitem{hoboya}  C. M. Ho, D. Boyanovsky, H. J. de Vega, Phys.Rev.
\textbf{D72}  085016 (2005).

\bibitem{lee} B. W. Lee and R. Shrock, Phys. Rev. \textbf{D16},
1444,  1977.

\bibitem{petcov} S. T. Petcov, Yad. Fiz. \textbf{25}, 641 [Sov. J. Nucl.
Phys. \textbf{25}, 340] (1977).

\bibitem{bile} S. M. Bilenky and S. T. Petcov, Rev. of Mod. Phys.
\textbf{59}, 671 (1987).

\bibitem{WMAP}D. N. Spergel \textit{et al.}, Astrophys. J. Suppl.
\textbf{148}, 175 (2003).

\bibitem{hansen} S. H. Hansen, G. Mangano, A. Melchiorri, G Miele
and O. Pisanti, Phys. Rev. \textbf{D65}, 023511 (2001).

\bibitem{kneller} J. P. Kneller, R. J. Scherrer, G. Steigman and T.
P. Walker, Phys. Rev. \textbf{D64}, 123506 (2001).

\bibitem{luna} C. Lunardini and A. Yu. Smirnov, Phys. Rev.
\textbf{D64}, 073006 (2001).

\bibitem{doleq}A. D. Dolgov, S. H. Hansen, S. Pastor, S. T.
Petcov, G. G. Raffelt and D. V. Semikoz, Nucl. Phys. \textbf{B632},
363 (2002).

\bibitem{wong} Y. Y. Y. Wong, Phys. Rev. \textbf{D66}, 025015
(2002).

\bibitem{aba} K. Abazajian, J. F. Beacom and N. F. Bell, Phys.
Rev. \textbf{D66}013008 (2002).

\bibitem{entanglement} C. Giunti, hep-ph/0409230; A.D.Dolgov, A.Yu.Morozov,
L.B.Okun, M.G.Schepkin, Nucl.Phys. \textbf{B502}3 (1997); Y.
Srivastava, A. Widom and E. Sassaroli, Eur. Phys. J. \textbf{C2},
769 (1998); Y. N. Srivastava and A. Widom, hep-ph/9707268.


\bibitem{dol} A. Dolgov, Yad. Fiz. \textbf{33}, 1309 (1981)[Sov. J. Nucl.
Phys.\textbf{33}, 700 (1981)].

\bibitem{stod} L. Stodolsky, Phys. Rev. \textbf{D36}, 2273 (1987); R. A. Harris,
L. Stodolsky, Phys. Lett. \textbf{B116}, 464 (1982); R. A. Harris
and L. Stodolsky, Phys. Lett. \textbf{B78}, 313 (1978).

\bibitem{mano} A. Manohar, Phys. Lett. \textbf{B186}, 370 (1987).

\bibitem{slichter} C. P. Slichter, \textit{Principles of Magnetic
Resonance}, (Springer-Verlag, Berlin, Heidelberg, 1978).

\bibitem{rafstod} G. Raffelt, G. Sigl and L. Stodolsky, Phys. Rev.
Lett. \textbf{70}, 2363 (1993); Phys. Rev. \textbf{D45}, 1782
(1992).

\bibitem{rafsigl} G. Sigl and G. Raffelt, Nucl. Phys. \textbf{B406},
423 (1993).

\bibitem{mckellar} B. H. J. McKellar and M. J. Thomson, Phys. Rev.
\textbf{D49}, 2710 (1994).

\bibitem{giunti} C. Giunti, Eur.Phys.J. \textbf{C39 }, 377 (2005).

\bibitem{blasone} M. Blasone and  G. Vitiello, Annals Phys. \textbf{244}, 283
(1995); Phys.Rev. \textbf{D60} (1999) 111302; E. Alfinito, M.
Blasone, A. Iorio, G. Vitiello, Phys.Lett. \textbf{B362}, 91 (1995).

\bibitem{fujii} K. Fujii, C. Habe, T. Yabuki, Phys.Rev. \textbf{D59}
113003 (1999); Phys.Rev. \textbf{D64}, 013011 (2001).

\bibitem{field} J. H. Field, hep-ph/0303241; hep-ph/0503034.

\bibitem{ji} C-R. Ji and Y. Mishchenko, Phys. Rev. \textbf{D65},
 096015 (2002); Annals Phys. \textbf{315} 488 (2005).

\bibitem{li} Y. F. Li and Q. Y. Liu, hep-ph/0604069.

\bibitem{hoboydense}  D. Boyanovsky, C.\, M. Ho, Phys.Rev. \textbf{D69},125012
(2004).



\bibitem{fullerQZ} K. Abazajian, G.M. Fuller, M. Patel, Phys.Rev. \textbf{D64}
(2001) 023501.

\bibitem{footQZ} R. Foot and R. R. Volkas, Phys. Rev. \textbf{D55},
5147 (1997).

\bibitem{htl} E. Braaten and R. Pisarski, Nucl. Phys.
\textbf{B337},569 (1990); \textbf{B339},310 (1990); R. Pisarski,
Phys. Rev. Lett. \textbf{63}, 1129 (1989); Nucl. Phys.
\textbf{A525}, 175 (1991).

\bibitem{bellac} M. Le Bellac, \textit{Thermal Field Theory},  (Cambridge
University Press, Cambridge, England, 1996).

\bibitem{iancu} J.-P. Blaizot and E. Iancu, Phys. Rev. Lett.
\textbf{76}, 3080 (1996); Phys. Rev. \textbf{D55}, 973 (1997);
\textbf{56}, 7877  (1997).

\bibitem{anomadamboya} S.-Y. Wang, D. Boyanovsky, H. J. de Vega
and D.-S. Lee, Phys. Rev. \textbf{D62}, 105026 (2000); D.
Boyanovsky, H. J. de Vega, S.-Y. Wang, Phys.Rev. \textbf{D67},
065022 (2003);  D. Boyanovsky, H. J. de Vega,  Annals Phys.
\textbf{307}, 335 (2003).

\bibitem{kamland} KamLAND Collaboration, K. Eguchi \textit{et al.},
 Phys. Rev. Lett., \textbf{90}, 021802 (2003); T. Araki \textit{et al.},
hep-ex/0406035.







\end{thebibliography}
\end{document}